\documentclass[%
 reprint,
 twocolumn,
 amsmath,amssymb,
 aps,
 prapplied,
 superscriptaddress
]{revtex4-2}

\usepackage{graphicx}
\usepackage{dcolumn}
\usepackage{bm}
\usepackage{braket}
\usepackage[dvipsnames, x11names]{xcolor}
\usepackage[colorlinks=true]{hyperref}
\hypersetup{
    colorlinks=true,
    citecolor=blue,
    linkcolor=red,
    urlcolor=RoyalBlue,
    anchorcolor=Purple
}

\begin{document}


\title{\textbf{High-Resolution Dynamical Eigenspectroscopy via Variational Trotter Compression on a Superconducting Qubit Processor}
}%

\author{Liyang Sui}
\thanks{These authors contributed equally to this work.}
\affiliation{Zhejiang Province Key Laboratory of Quantum Technology and Device, School of Physics, Zhejiang University, Hangzhou 310027, China}
\author{Xingrui Liu}
\thanks{These authors contributed equally to this work.}
\affiliation{Zhejiang Province Key Laboratory of Quantum Technology and Device, School of Physics, Zhejiang University, Hangzhou 310027, China}
\author{Yufan Li}
\thanks{These authors contributed equally to this work.}
\affiliation{Zhejiang Province Key Laboratory of Quantum Technology and Device, School of Physics, Zhejiang University, Hangzhou 310027, China}
\author{Sainan Huai}
\affiliation{Tencent Quantum Laboratory, Tencent, Shenzhen 518057, China}

\author{Zhiwen Zong}
\affiliation{Tencent Quantum Laboratory, Tencent, Shenzhen 518057, China}
\author{Kunliang Bu}
\affiliation{Tencent Quantum Laboratory, Tencent, Shenzhen 518057, China}
\author{Xiaopei Yang}
\affiliation{Tencent Quantum Laboratory, Tencent, Shenzhen 518057, China}
\author{Wenyan Jin}
\affiliation{Zhejiang Province Key Laboratory of Quantum Technology and Device, School of Physics, Zhejiang University, Hangzhou 310027, China}
\author{Bowen Chen}
\affiliation{Zhejiang Province Key Laboratory of Quantum Technology and Device, School of Physics, Zhejiang University, Hangzhou 310027, China}
\author{Xutao Zhang}
\affiliation{Zhejiang Province Key Laboratory of Quantum Technology and Device, School of Physics, Zhejiang University, Hangzhou 310027, China}
\author{Jianlan Wu}\thanks{jianlanwu@zju.edu.cn}
\affiliation{Zhejiang Province Key Laboratory of Quantum Technology and Device, School of Physics, Zhejiang University, Hangzhou 310027, China}

\author{Shengyu Zhang} \thanks{shengyzhang@tencent.com}
\affiliation{Tencent Quantum Laboratory, Tencent, Shenzhen 518057, China}

\author{Yi Yin} \thanks{yiyin@zju.edu.cn}
\affiliation{Zhejiang Province Key Laboratory of Quantum Technology and Device, School of Physics, Zhejiang University, Hangzhou 310027, China}
\date{\today}

\begin{abstract}
The pursuit of high-resolution eigenspectroscopy on noisy intermediate-scale
quantum devices is often hindered by the trade-off between circuit
depth and coherence time. In this work, we introduce and experimentally
demonstrate a dynamical eigenspectroscopy protocol that extracts
fine-grained energy structures from time-dependent survival amplitudes.
To overcome the finite coherence window of current superconducting processors,
we employ Variational Trotter Compression (VTC)
as a practical means to extend the duration of high-fidelity unitary
evolution. Using a multi-connected 9-qubit superconducting processor,
we reconstruct the time-domain autocorrelation signal $\langle\psi_0|\psi(t)\rangle$
via quantum state tomography for the $\mathrm{H_2}$ molecule at different
bond lengths and for the Fermi-Hubbard model across different correlation regimes.
Through multi-frequency fitting and Fourier analysis,
the extracted eigenenergies agree with the exact-diagonalization values to within
$2\times10^{-3}$, including the near-degenerate levels in the strongly interacting regime.
Our results establish experimental dynamical spectroscopy as a robust and
generalizable framework for simulating both quantum chemistry and strongly correlated lattice systems,
bridging weak- and strong-coupling regimes on near-term quantum hardware.
\end{abstract}

\maketitle

\section{Introduction}

Quantum simulation represents a flagship application of
noisy intermediate-scale quantum (NISQ)
processors~\cite{Lloyd1996UniversalQuantumSimulators,AspuruGuzik2005SimulatedQuantum,
Arute2019QuantumSupremacy,Arute2020HartreeFock,Kjaergaard2020SuperconductingQubits}.
Digital approaches, which decompose the time-evolution
operator into universal gate sequences, have successfully
tackled diverse problems ranging from molecular electronic
structure---exemplified by $\mathrm{H_2}$ energy-surface
reconstructions on superconducting
qubits~\cite{OMalley2016ScalableMolecular,Colless2018MolecularSpectra,
PhysRevApplied.16.034050,Zong2024MolecularEnergies}---to
the dynamics of many-body condensed-matter systems
~\cite{Hu2025SpontaneousSymmetryBreaking,Andersen2025ThermalizationCriticality,
Liu2026Prethermalization,Fischer2026ManyBodyQuantumChaos}.
To mitigate hardware noise, the variational quantum eigensolver (VQE)
has emerged as a leading paradigm for preparing ground and
low-lying excited states with shallow circuits~\cite{Kandala2017HardwareEfficientVQE,
Higgott2019VariationalExcited,Gocho2023ExcitedVQE,Zhang2021AdaptiveVQE}.
Recent hardware demonstrations have extended VQE to the
simultaneous characterization of multiple eigenstates~\cite{
Xu2023ConcurrentQuantumEigensolver,Zhang2026SimultaneousEnergyLevels}.
In our prior work on the multi-connected 9-qubit L9 processor, we
employed an extended VQE with a phase-based loss function
to digitally simulate 3D Ising criticality~\cite{Sui2026_3DIsingPlatonic}.
While this approach enabled high-resolution eigenenergy
extraction via Fourier analysis of classically computed
time series, it relied on post-processed classical evolution
rather than genuine in-situ quantum dynamics. Bridging
this gap---realizing authentic on-chip time evolution for
eigenstate preparation and spectroscopy---remains
a pivotal challenge for quantum simulation.

To address the circuit depth limitations of standard Trotter
evolution~\cite{Salathe2015DigitalSpinModels,Barends2015DigitalFermionicModels},
we implement the established Variational Trotter Compression (VTC)
protocol~\cite{Berthusen2022VariationalTrotter}. This hybrid scheme
interleaves short Trotter steps with variational re-compression
into a shallow, hardware-efficient ansatz, thereby maintaining
high-fidelity overlap with the ideal evolution trajectory.
Central to achieving high-resolution dynamical eigenspectroscopy
is the reconstruction of the survival amplitude
$S(t)=\langle\psi_0|\psi(t)\rangle$.
Since the relative phase information encoded in VTC parameters
is not directly accessible via measurement, we perform quantum
state tomography (QST) on the evolved state at discrete time
points and then align these experimental data to the numerically
simulated phase trajectory via classical post-processing. This
hybrid classical-quantum workflow yields a continuous
time series suitable for spectral analysis. By avoiding the
deep multi-controlled gates required for ancilla-based phase
estimation~\cite{Somma2019EigenvalueTS}, our approach leverages
the shallow-circuit nature of VTC to suppress the cumulative
Trotter error that would otherwise render long-time evolution
infeasible, thereby accessing effective dynamics far exceeding
the bounds imposed by fixed-depth Trotter schemes.
The Fourier spectrum of $S(t)$ reveals the complete
eigenspectrum of example systems, demonstrating a robust
workflow for in-hardware quantum simulation.

Although the $\mathrm{H_2}$ molecule and the Fermi-Hubbard model
originate from quantum chemistry and condensed-matter physics
respectively, they are intimately linked through the interplay
of kinetic delocalization and interaction-driven localization.
At equilibrium, $\mathrm{H_2}$ is well described by single-reference
methods in which dynamic correlation dominates; upon bond stretching,
static correlation emerges as the two electrons tend to localize
on separate nuclei, necessitating a multi-reference description.
In the minimal STO-3G basis, $\mathrm{H_2}$ maps onto an effective
two-site Hubbard dimer: the hopping amplitude $t$ decays exponentially
with bond length $R$, while the on-site interaction $U$ remains
approximately constant. Consequently, increasing $R$ drives the
ratio $U/t$ from the weak- to the strong-coupling regime, making
$\mathrm{H_2}$ dissociation a molecular prototype of the Mott crossover.
Extending beyond the dimer, we conducted experiments on L9, a 9-qubit
superconducting processor, and used four-qubit subsets of the processor to
realize a finite impurity representation of the single-band Hubbard model
obtained within two-site dynamical mean-field theory (DMFT)
~\cite{Keen2020TwoSiteDMFT,Stanisic2022FermiHubbardGroundState,Nie2024SelfConsistentSIAM}.
The investigated parameter points cover an effective progression
from weak to strong correlation, ranging from the noninteracting
limit to a strongly correlated, Mott-like regime.
Our VTC results on both model settings reinforce this connection:
for $\mathrm{H_2}$ at five equally
spaced bond lengths from $R=0.25\,\mathrm{\AA}$ to
$R=2.25\,\mathrm{\AA}$, the average fidelity between
the experimentally evolved and numerically simulated
states exceeds $99.6\%$ over the sampled evolution times,
and the extracted eigenenergies agree with exact
diagonalization to within $2\times10^{-3}$ across all
investigated bond lengths; for the single-band Hubbard model,
the same VTC ansatz with depth $L\geq2$ suppresses
the evolution error below $10^{-4}$ across extended time
scales. This high-resolution spectral retrieval is enabled by
the extended evolution windows afforded by VTC, which provide
the frequency resolution necessary to resolve closely
spaced eigenvalues. Collectively, these results establish
VTC as a unified hardware-efficient framework capable of
describing the crossover from weak to strong correlation,
and bridge the quantum chemistry $\mathrm{H_2}$ molecule
and the condensed-matter Hubbard model. We note that while
true Mott phase transitions require the thermodynamic limit,
our finite-size simulation faithfully captures this essential
competition between kinetic delocalization and
interaction-driven localization.

\medskip

\section{Theoretical Framework and Numerical Simulation}
\label{sec:theory}

We adopt natural units throughout this work, with $\hbar = 1$.
Consider an arbitrary time-independent Hamiltonian $H = \sum_\alpha H_\alpha$,
where the constituent terms $\{H_\alpha\}$ are mutually non-commuting.
For any initial state $\ket{\psi_0}$, its expansion in the
eigenbasis $\{\ket{\phi_i}\}$ of $H$ reads
\begin{equation}\label{eq:expansion}
\ket{\psi_0} = \sum_i c_i \ket{\phi_i}, \quad c_i = \bra{\phi_i}{\psi_0}\rangle.
\end{equation}
Under unitary evolution governed by $H$, the time-evolution operator
$U(t) = e^{-iHt}$ yields
\begin{equation}\label{eq:evolution}
U(t)\ket{\psi_0} = \sum_i c_i\, e^{-iE_i t}\ket{\phi_i},
\end{equation}
where $E_i$ denotes the eigenenergy associated with $\ket{\phi_i}$.
Taking the expectation value of $U(t)$ with respect to $\ket{\psi_0}$
then gives
\begin{equation}\label{eq:signal}
\bra{\psi_0}U(t)\ket{\psi_0}= \sum_i |c_i|^2\, e^{-iE_i t}.
\end{equation}
This signal constitutes a coherent superposition of oscillatory components,
each oscillating at frequency $E_i$ (in natural units); the amplitude of
the $i$-th component is precisely $|c_i|^2$, and thus the $i$-th eigenenergy
contributes measurably to the signal only if $|c_i|^2 > 0$.
Accordingly, multi-frequency fitting and Fourier spectral analysis of the
time-dependent signal in Eq.~\eqref{eq:signal} enable the extraction of
all eigenenergies $\{E_i\}$ for which the initial state exhibits
non-vanishing overlap.

\begin{figure}[t]
\centering
\includegraphics[width=1.0\columnwidth]{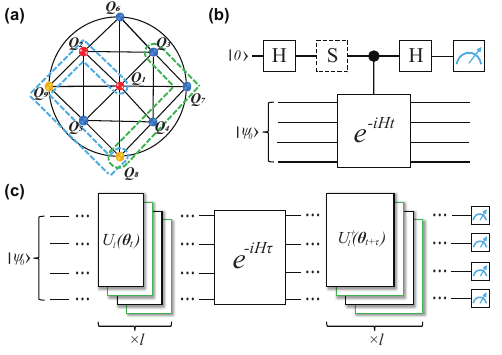}
\caption{Device schematic and quantum circuits.
(a) Topology of the L9 processor. Central qubit $Q_1$ links
to peripherals $Q_2$--$Q_9$ via tunable couplers (total: 24). Highlighted
subsets: red/yellow circles ($Q_1,Q_2$; $Q_8,Q_9$) for $\mathrm{H_2}$ simulation;
blue/green boxes ($Q_1,Q_2,Q_8,Q_9$; $Q_3,Q_4,Q_7,Q_8$) for Hubbard model.
(b) Hadamard test circuit with ancilla (top). Sequential $H$, $S$,
controlled--$e^{-iHt}$, and $H$ gates precede measurement.
Controlled dynamic evolution introduce significant overhead.
(c)  Ancilla-free Variational Trotter Compression (VTC) protocol for dynamic evolution:
Given an initial state $\ket{\psi_0}$, the hardware-efficient ansatz
$U(\boldsymbol{\theta}_t)$---pre-optimized at time $t$ is first applied ($\ell$: layers;
black/green: odd/even layers with swapped CNOT order).
A short-time evolution $e^{-iH\tau}$ is then enacted, followed by
$U^\dagger(\boldsymbol{\theta}_{t+\tau})$. Minimizing the infidelity
between the output state and $\ket{\psi_0}$ enables gradient-based
optimization of $\boldsymbol{\theta}_{t+\tau}$, such that
$U(\boldsymbol{\theta}_{t+\tau})\ket{\psi_0}$ approximates the
target time-evolved state at $t+\tau$.
}
    \label{fig1}
\end{figure}

To measure the expectation value $\bra{\psi_0}U(t)\ket{\psi_0}$,
the ancilla-assisted Hadamard test is a standard quantum circuit
primitive. Figure~\ref{fig1}(b) illustrates its operational principle:
an ancilla qubit is entangled with the system register via a
controlled-$U(t)$ gate. Starting from the initial joint state
$|\psi_{\text{tot}}\rangle = |0\rangle_a \otimes |\psi_0\rangle_s$,
application of a Hadamard gate to the ancilla prepares the
superposition $\frac{1}{\sqrt{2}}\big(|0\rangle_a + |1\rangle_a\big) \otimes |\psi_0\rangle_s$.
When a phase gate $S = \mathrm{diag}(1,i)$ is inserted before
the controlled-$U(t)$, it imparts a relative phase $i$ to
the $|1\rangle_a$ branch, yielding $\frac{1}{\sqrt{2}}\big(|0\rangle_a + i|1\rangle_a\big) \otimes |\psi_0\rangle_s$.
The subsequent controlled-$U(t)$ then transforms the state
into $\frac{1}{\sqrt{2}}\big(|0\rangle_a \otimes |\psi_0\rangle_s + i|1\rangle_a \otimes U(t)|\psi_0\rangle_s\big)$.
A final Hadamard gate on the ancilla maps the complex phase
information onto measurable population imbalances: measuring
the ancilla Pauli--$X$ expectation value $\langle X \rangle_a = P_0-P_1$---where $P_0$ ($P_1$)
denotes the probability of observing $|0\rangle_a$ ($|1\rangle_a$)---yields
$\operatorname{Re}\big[\bra{\psi_0}U(t)\ket{\psi_0}\big]$
in the absence of the $S$ gate, and
$\operatorname{Im}\big[\bra{\psi_0}U(t)\ket{\psi_0}\big]$
when the $S$ gate is included.
However, for practical experiments, the dynamical evolution operator
$U(t)$ is typically implemented via a Trotterized decomposition
in digital quantum algorithms, introducing significant circuit depth.
The controlled-$U(t)$ operation further increases gate complexity,
notably amplifying the CNOT gate count~\cite{Yang2024PhaseSensitive}.

Here, we adopt an ancilla-free protocol to measure the expectation
value $\bra{\psi_0}U(t)\ket{\psi_0}$. For experimental simplicity,
the initial state is selected as a computational basis state---specifically,
a product state amenable to high-fidelity preparation.
As a representative two-qubit instance, we choose
$\ket{\psi_0} = \ket{00}$. Under this choice, Eq.~\eqref{eq:signal}
simplifies to the $(1,1)$ matrix element of $U(t)$, i.e., the
amplitude of $U(t)\ket{00}$ in the $\ket{00}$ basis state---where
the labeling convention identifies $\ket{00}$ with the first standard
basis vector in the $2^n$-dimensional Hilbert space:
\begin{equation}\label{eq:element}
\bra{00}U(t)\ket{00} = \bigl[U(t)\ket{00}\bigr]_1.
\end{equation}
By fitting the experimentally acquired time-series data of
this element and performing Fourier analysis,
the system eigenenergies are extracted directly---eliminating the
need for ancilla qubits or controlled-unitary gates.

Define a short time step $\tau$, a total evolution time $t$
is partitioned into $N = t/\tau$ discrete
intervals.
For sufficiently small $\tau$, the time evolution over a single
step is well approximated by the first-order Lie-Trotter
product formula:
\begin{equation}\label{eq:trotter}
S_1(\tau) \equiv \prod_{\alpha} e^{-iH_\alpha\tau},
\quad \tau = t/N.
\end{equation}
Each factor $e^{-iH_\alpha\tau}$ can be implemented as a quantum circuit
composed of single- and two-qubit gates, provided that $H_\alpha$
admits a Pauli-operator decomposition.
Since the Hamiltonian terms $\{H_\alpha\}$ are generally non-commuting,
the $N$-fold composition $\bigl(S_1(t/N)\bigr)^{N}$ then approximates the
exact unitary $e^{-iHt}$. A rigorous error bound in the spectral norm follows from
the Baker-Campbell-Hausdorff (BCH) expansion and application of
the triangle inequality across all $N$ Trotter steps.
Specifically, the additive approximation error satisfies
\begin{equation}\label{eq:trotter-error}
\begin{split}
\Bigl\|\bigl(S_1(t/N)\bigr)^{N} - e^{-iHt}\Bigr\|
&\leq \frac{t^2}{2N}\sum_{\alpha<\beta}
\bigl\|[H_\alpha,H_\beta]\bigr\| \\
&\quad + \mathcal{O}(t^3/N^2),
\end{split}
\end{equation}
where $\|\cdot\|$ denotes the spectral norm and the sum extends
over all unordered pairs $(\alpha,\beta)$ for which
$[H_\alpha,H_\beta] \neq 0$. This bound exhibits commutator scaling~\cite{Childs2021TrotterError}:
the leading-order error is proportional to the sum of the spectral
norms of the pairwise commutators. For a fixed Trotter step $\tau$ ($=t/N$),
the simulation error scales linearly with the evolution time $t$.
Consequently, achieving a target precision $\varepsilon$ necessitates
a circuit depth proportional to $t^2$. However, practical quantum
processors face inherent constraints from finite coherence times
and gate infidelities. This fundamental limitation underscores
the critical need for strategies to mitigate circuit depth in
long-time dynamics.

The VTC method
addresses this challenge by substituting the Trotter circuit at each
time $t$ with a shallow-depth, parameterized quantum ansatz~\cite{Berthusen2022VariationalTrotter}.
As illustrated in Fig.~\ref{fig1}(c), the algorithm proceeds as follows:
For each step $m = 1, 2, \dots, N$, an ansatz $U(\boldsymbol{\theta}_m)$
is iteratively constructed to approximate the
cumulative unitary $e^{-iH m\tau}$. Specifically, let
$\boldsymbol{\theta}_{m-1}^{\mathrm{opt}}$ denote the optimal
parameters obtained at step $m-1$, so that $U(\boldsymbol{\theta}_{m-1}^{\mathrm{opt}})$
approximates $e^{-iH(m-1)\tau}$. Given an initial state $\ket{\psi_0}$,
first apply $U(\boldsymbol{\theta}_{m-1}^{\mathrm{opt}})$ to obtain
the intermediate state $U(\boldsymbol{\theta}_{m-1}^{\mathrm{opt}})\ket{\psi_0}$;
then apply the exact short-time evolution $e^{-iH\tau}$ to yield the target
state for step $m$:
\begin{equation}\label{eq:target_m}
\ket{\phi_m^{\mathrm{target}}} = e^{-iH\tau} U(\boldsymbol{\theta}_{m-1}^{\mathrm{opt}}) \ket{\psi_0}.
\end{equation}
Next, apply the adjoint ansatz $U^\dagger(\boldsymbol{\theta}_m)$ to
this target state, producing the measurable final state
$U^\dagger(\boldsymbol{\theta}_m)\ket{\phi_m^{\mathrm{target}}}$. The
objective is to drive this final state toward $\ket{\psi_0}$---i.e., to
implement an effective identity operation over the full step. Relative
to the original Berthusen \textit{et al.} approach~\cite{Berthusen2022VariationalTrotter},
we adopt the $\ell^1$ norm (Manhattan distance) as the optimization objective instead of
the $\ell^2$ norm (Euclidean distance), aligning more naturally with
experimentally reconstructed probability amplitudes. Accordingly,
the loss function is defined as the $\ell^1$ norm of the difference between
the final and initial states $\bm{\delta}(\boldsymbol{\theta}_m)$:
\begin{equation}\label{eq:manhattan}
\begin{split}
\mathcal{L}(\boldsymbol{\theta}_m)
&= \bigl\|\,\bm{\delta}(\boldsymbol{\theta}_m)\,\bigr\|_1
 = \sum_{j=1}^{2^n} \bigl|\,\delta_j(\boldsymbol{\theta}_m)\,\bigr|, \\
\bm{\delta}(\boldsymbol{\theta}_m)
&\equiv \ket{\psi_0}
  - U^{\dagger}(\boldsymbol{\theta}_m)\,e^{-iH\tau}\,U(\boldsymbol{\theta}_{m-1}^{\mathrm{opt}})\ket{\psi_0},
\end{split}
\end{equation}
in which the index $j$ spans all $2^n$ basis states
$\{\ket{k}\}_{k=1}^{2^n}$, and the $\ell^1$ norm is defined as
$\|\bm{\delta}(\boldsymbol{\theta})\|_1 = \sum_{j=1}^{2^n} |\delta_j(\boldsymbol{\theta})|$.
The optimal parameter vector
$\boldsymbol{\theta}_m^{\mathrm{opt}} = \arg\min_{\boldsymbol{\theta}}\,\mathcal{L}(\boldsymbol{\theta})$
yields a circuit $U(\boldsymbol{\theta}_m^{\mathrm{opt}})$ that
approximates the cumulative time-evolution operator $e^{-iH m\tau}$ to high fidelity.

\begin{figure*}[tp]
    \centering
    \includegraphics[width=0.8\linewidth]{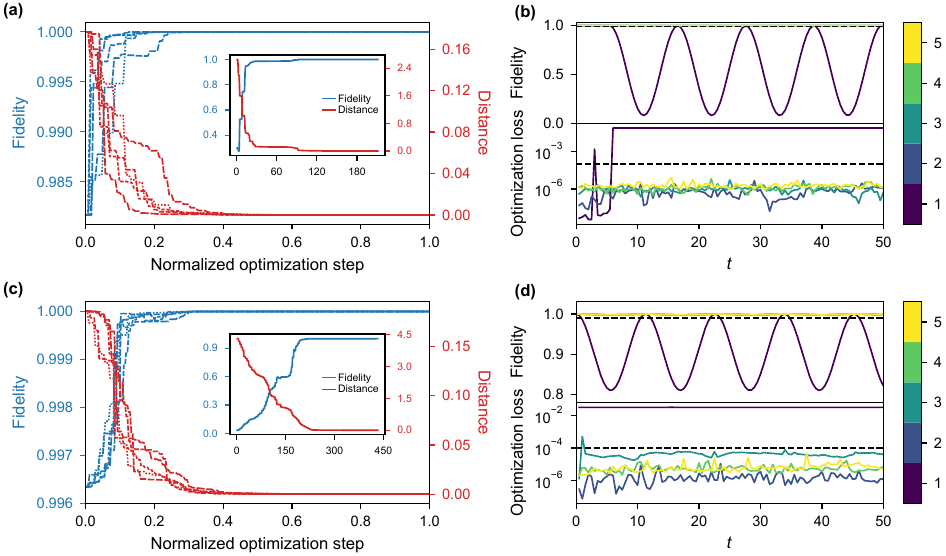}
    \caption{VTC-optimized numerical simulations.
    The top row (a), (b) corresponds to the hydrogen
    molecule ($R = 2.25\,\mathrm{\AA}$, initial state $\ket{00}$), while
    the bottom row (c), (d) shows the Hubbard intermediate
    state ($U=1$, $V=0.121$, initial state $\ket{0100}$).
    (a), (c) Convergence of the L-BFGS-B optimizer.
    Different lines represent results at different evolution time ($t=10, 20, 30, 40, 50$);
    the inset highlights slow initial convergence at $t=0.5$ due to cold-starting.
    Subsequent steps employ a warm-start, converging with fewer steps(blue: fidelity
    rising to 1; red: Manhattan distance decreasing to 0; $x$-axis: normalized optimization step number).
    (b), (d) Scan over ansatz depth ($L=1$--$5$, color-coded on the right):
    The upper panel shows the evolution state fidelity, while the lower panel displays
    the $\ell^1$ optimization loss as functions of evolution time $t$.
    For $L=1$, the fidelity cannot maintain high-precision evolution over time;
    for $L\geq 2$, the loss can be stably suppressed below $10^{-4}$ (black dashed line),
    enabling the fidelity to remain above 0.99---this criterion guides
    the selection of the ansatz depth.
    }
    \label{fig2}
\end{figure*}

For the $m$-th optimization step, the initial parameter vector is
set to the solution from the preceding step:
$\boldsymbol{\theta}_m^{(0)} = \boldsymbol{\theta}_{m-1}^{\mathrm{opt}}$.
This warm-start strategy exploits the continuity of quantum dynamics---since
the state evolution over a single Trotter interval $\tau$ is infinitesimal,
the optimal parameters for adjacent steps are highly correlated. Consequently,
convergence is substantially accelerated. The warm-start strategy is similar to
the principle of adiabatic parameter initialization used in prior variational
adiabatic eigensolver~\cite{Liu2026SiliconBandStructure}.
At each evolution step, $\boldsymbol{\theta}$ is optimized by minimizing
the loss function $\mathcal{L}$ with the gradient-based quasi-Newton
L-BFGS-B algorithm~\cite{Byrd1995LBFGSB}. Numerical gradients are employed, and the optimizer
is configured with a tolerance of $10^{-16}$ in the noiseless simulations.

Prior to experimental deployment on superconducting quantum processors,
we conduct comprehensive numerical benchmarking of the VTC framework
across two paradigmatic quantum many-body systems: the hydrogen
molecule (2-qubit system) and the strongly correlated Fermi-Hubbard
model (4-qubit system). We systematically characterize both the
convergence properties and the minimal ansatz depth required to
achieve target fidelity. Results are summarized in Fig.~\ref{fig2}.
Figure \ref{fig2}(a) and \ref{fig2}(b) (top row) present data for the
hydrogen molecule at bond length $R = 2.25\,\text{\AA}$, initialized
in $\ket{00}$, with Trotter step $\tau = 0.5$; figure \ref{fig2}(c) and \ref{fig2}(d) (bottom row)
correspond to an intermediate state of the Hubbard model with
on-site interaction $U = 1$, nearest-neighbor hopping $V = 0.121$,
initial state $\ket{0100}$, $\tau = 0.5$, and total
evolution steps $N = 100$.

Crucially, VTC does not aim to replicate the full unitary $U(t)$
but instead targets the output state $\ket{\psi(t)}$ directly.
VTC ensures that the time-domain
signal $\bra{\psi_0}U(t)\ket{\psi_0}$ reproduced with high accuracy
relative to the ideal evolution, and can be viewed
as a resource-efficient approximation to the time-evolution.
It substantially reduces the circuit depth while retaining the relevant
dynamical observable, without altering the underlying Hamiltonian
or introducing a new dynamical model.

The fidelity and distance shown in Fig.~\ref{fig2}
quantify the approximation quality of the variational state to the
exact time-evolved state at each Trotter step. Specifically,
let $\ket{\psi^{\mathrm{V}}_m} = U(\boldsymbol{\theta}_m)\ket{\psi_0}$
denote the variational state obtained after optimizing the ansatz
at the $m$-th step, and let $\ket{\varphi_m} = e^{-iH m\tau}\ket{\psi_0}$
be the corresponding exact evolved state under the full Hamiltonian.
The fidelity is defined as the squared modulus of their quantum
mechanical overlap:
\begin{equation}
\label{eq:sim-fidelity}
f(\boldsymbol{\theta}_m) = \bigl|\langle{\varphi_m}|{\psi^{\mathrm{V}}_m}\rangle\bigr|^2.
\end{equation}
The distance is defined as the $\ell^1$  norm---also known as the
Manhattan distance---of the difference between their vector
representations in the computational basis:
\begin{equation}\label{eq:sim-distance}
D(\boldsymbol{\theta}_m) = \bigl\|\,\ket{\varphi_m} - \ket{\psi^{\mathrm{V}}_m}\,\bigr\|_1 = \sum_{j=1}^{2^n} \bigl|(\varphi_m)_j - (\psi^{\mathrm{V}}_m)_j\bigr|.
\end{equation}
Note that the optimization procedure does not rely on prior knowledge of $\ket{\varphi_m}$.
The two metrics provide complementary characterizations of
approximation accuracy: fidelity captures global phase-insensitive
similarity (i.e., closeness as quantum states), whereas $\ell^1$
distance quantifies local, component-wise deviation in vector space.
Consequently, high fidelity ($f \to 1$) and low distance ($D \to 0$)
jointly indicate high-fidelity VTC compression.

Figure~\ref{fig2}(a) and \ref{fig2}(c) depict the convergence behavior of
the inner-layer L-BFGS-B optimization for the hydrogen molecule
and the Fermi-Hubbard model, respectively. Each solid or dashed line
corresponds to a distinct Trotter time step $m\tau$, with the horizontal
axis representing normalized iteration counts. At the initial
step ($t = \tau = 0.5$), the fidelity converges slowly due to poor
initialization far from the optimal solution (inset). In contrast,
subsequent time steps exhibit faster convergence,
enabled by a warm-start strategy that leverages optimized parameters
from preceding steps to minimize the distance between adjacent quantum states.
For $\mathrm{H_2}$ ($L=2$, 8 params), the $\ell^1$ loss plateaus at
$\sim\!10^{-6}$ within 200 steps, yielding fidelities $>\!99.9\%$
and collapsing the state distance from $\sim\!0.18$ to $10^{-6}$.
Likewise, in the intermediate Hubbard regime ($L=2$, 16 params),
the optimizer converges to a comparable loss of $\sim\!10^{-6}$
within just 100 steps, while sustaining fidelities above $>\!99.9\%$.
These results confirm that VTC robustly steers variational states
toward their exact time-evolved counterparts with high fidelity
and controlled amplitude error.

Figure~\ref{fig2}(b) and \ref{fig2}(d) present layer-sweep
results ($L=1$--$5$, color bar right) for both systems. The
upper panels plot the state fidelity, while the lower panels show
the $\ell^1$ optimization loss versus evolution time $t$. A single
layer ($L=1$) lacks the expressive power to suppress the loss below
the target threshold, causing fidelity to degrade markedly with
increasing $t$. In contrast, $L \geq 2$ stabilizes the loss below
$10^{-4}$ (black dashed line). This threshold suffices to sustain
VTC fidelity above $99\%$ (dashed line) across all time steps.
Accordingly, we use the shallowest ansatz depth satisfying this
criterion throughout the main text: $L=2$ for the $\mathrm{H_2}$
molecule and $L=2$ or $3$ for the Hubbard model.

\section{Experimental Results}
\label{sec:experiment}

Following the numerical results, experiments were performed on the L9 superconducting quantum processor~\cite{Sui2026_3DIsingPlatonic}, a flip-chip-integrated device featuring nine frequency-tunable transmon
qubits ($Q_1$--$Q_9$)~\cite{Barends2013CoherentJosephsonQubit} and 24 independently controlled Josephson-based tunable
couplers (connectivity depicted in Fig.~\ref{fig1}(a))~\cite{Sete2021FloatingTunableCoupler}.
The flip-chip processor is fabricated on two sapphire substrates: the lower substrate
hosts microwave control lines and high-coherence coplanar waveguide resonators
for dispersive readout, while the upper substrate integrates the qubits
and their associated tunable couplers~\cite{Bu2025TantalumAirbridges}. Crucially, the floating tunable
coupler architecture enables precise dynamic suppression of
residual $ZZ$ crosstalk~\cite{Sete2021FloatingTunableCoupler,Li2020TunableCouplerCZ}.
For $\mathrm{H_2}$ simulations,
we utilized two disjoint pairs---($Q_1$, $Q_2$) and ($Q_8$, $Q_9$), highlighted in
red and yellow in Fig.~\ref{fig1}(a), respectively. For the Fermi-Hubbard model,
we programmed two distinct four-qubit sublattices: {$Q_1$, $Q_2$, $Q_8$, $Q_9$} (blue
dashed box) and {$Q_1$, $Q_3$, $Q_7$, $Q_8$} (green dashed box), activating native
couplings as required. Leveraging the high connectivity of the L9 processor, we can implement
in-situ-calibrated, high-fidelity two-qubit gates on arbitrary qubit pairs
allowed by the connectivity, thereby enabling more flexible design of the
hardware-efficient ansatz circuits.

Each transmon has a set frequency in the range of $4$-$5$ GHz and an anharmonicity of
approximately $-200$ MHz, with an average energy relaxation time
$T_1\!=\!57.3\,\mu\text{s}$ ($37.9-85.6\,\mathrm{\mu s}$) and an average
spin-echo dephasing time $T_2^{\mathrm{echo}}\!=\!26.8\,\mu\text{s}$ ($10.3-47.3\,\mathrm{\mu s}$).
Single-qubit rotation gates have an average fidelity of $99.93\%$ ($99.89\%-99.95\%$)
measured via Clifford randomized benchmarking, while two-qubit CZ gates
are implemented via floating tunable couplers with an average fidelity
of $99.46\%$ ($98.36\%-99.94\%$). The gate times---approximately 40 ns
for single-qubit and 30 ns for two-qubit operations---are significantly
shorter than both $T_1$ and $T_2^{\mathrm{echo}}$, effectively suppressing
decoherence errors. More detailed qubit parameters can be found in previous
literature~\cite{Sui2026_3DIsingPlatonic}.

Unlike prior work~\cite{Sui2026_3DIsingPlatonic,Liu2026SiliconBandStructure},
the final state in this experiment is
reconstructed as a density matrix $\rho$ via quantum state tomography (QST).
Prior to reconstruction, raw population measurements are corrected for
readout infidelity by applying the inverse of a pre-characterized assignment
error matrix $\mathbf{F}$. Full details regarding the choice of tomographic
measurement bases, the density matrix reconstruction, and the experimental
calibration protocol for $\mathbf{F}$ are provided in
Appendix~\ref{sec:qst-readout}. Following QST, we perform eigen-decomposition
on the reconstructed experimental density matrix $\rho^{\exp}$ and select
the eigenvector corresponding to the largest eigenvalue---thereby obtaining
the dominant pure-state component $\ket{\psi^{\exp}(t)}$. This purification
step mitigates mixedness arising from residual decoherence and measurement
imperfections.

Raw density matrices from QST lack sensitivity to the global
phase of the underlying quantum state, introducing time-random phase
offsets in the purified evolved state. Ancilla qubits resolve this
ambiguity by serving as an in-situ phase reference, as implemented
in the Hadamard test protocol. As detailed in Appendix~\ref{sec:phase},
an alternative pathway leverages the invariance of Hermitian
observable expectation values (e.g., Pauli strings) under global
phase rotations: one can directly compute the real-time evolution of
these observables from the density matrix or purified state to
extract energy level differences. For experimental simplicity,
however, we adopt a classical post-processing routine that
aligns the global phase of the QST-reconstructed purified
state to a numerically simulated reference trajectory at
ach time step (Appendix~\ref{sec:phase}). This approach effectively
removes phase-induced artifacts without incurring ancilla overhead.

\begin{figure*}[tp]
    \centering
    \includegraphics[width=0.8\linewidth]{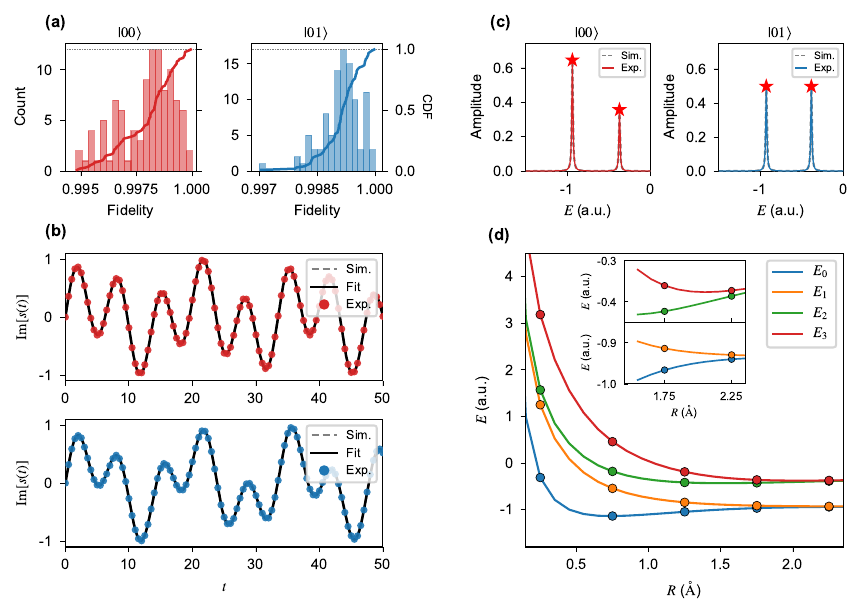}
    \caption{Experimental VTC dynamics and eigenspectra of $\mathrm{H_2}$ molecule.
    (a) Histogram (left axis, Count) and cumulative distribution
    function (right axis, CDF) of state fidelity between experimentally
    prepared and VTC-simulated states at each time step, evaluated for
    both $\ket{00}$ and $\ket{01}$ at the bond length $R = 2.25\,\mathrm{\AA}$.
    (b) Time evolution of the imaginary component $\mathrm{Im}[s(t)]$
    of the quantum state for initial states $\ket{00}$ and $\ket{01}$, as
    reconstructed from VTC-based experiments. Experimental data (Exp.) agree excellently
    with numerical simulations (Sim., gray dashes) across all time scales.
    Experimental data lead to multi-frequency sine fits (Fit, black lines),
    which are further Fourier-transformed for extraction of eigenspectra.
    (c) Fourier-transformed spectra of the time-resolved expectation
    values for the initial states $\ket{00}$ and $\ket{01}$, with high-resolution
    spectral peaks unambiguously identifying the system{\textquoteright}s eigenenergies.
    (d) Eigenenergy spectrum $E_0-E_3$ (data points) extracted
    via Fourier analysis of experimentally fitted time-domain signals at
    five bond distances $R$, compared against reference potential energy
    curves obtained from exact diagonalization of the molecular
    Hamiltonian (solid lines); inset highlights the near-degeneracy of
    excited states around $R = 1.75\,\mathrm{\AA}$ and $R = 2.25\,\mathrm{\AA}$.
    }
    \label{fig3}
\end{figure*}

The $\mathrm{H_2}$ molecule serves as a prototypical system for
covalent bond formation in the weakly correlated regime and,
conversely, as a canonical benchmark for probing the onset of
strong correlation at extended bond lengths. This dual role establishes
it as a cornerstone of quantum chemistry benchmarking.
We adopt the Hamiltonian form introduced in Ref.~\cite{Colless2018MolecularSpectra}
to describe the $\mathrm{H_2}$ molecule, which---being representable on
just two qubits---provides an ideal minimal model for quantum
simulation. The corresponding molecular Hamiltonian reads
\begin{equation}\label{eq:h2-ham}
H = \alpha_0\,\mathbf{I}
+ \alpha_1\,Z_1
+ \alpha_2\,Z_2
+ \alpha_3\,X_1 X_2
+ \alpha_4\,Z_1 Z_2,
\end{equation}
where the coefficients $\alpha_i$ are functions of the internuclear
bond distance $R$~\cite{Colless2018MolecularSpectra,PhysRevApplied.16.034050,Zong2024MolecularEnergies}.
Within the STO-3G minimal basis set, $\alpha_0$
denotes the weight of the identity operator (introducing only a
global phase shift), $\alpha_1$ and $\alpha_2$ parameterize the
local $Z$-basis terms acting on qubits 1 and 2, respectively,
$\alpha_3$ governs the $X_1X_2$ exchange interaction, and $\alpha_4$
quantifies the $Z_1Z_2$ two-body correlation term.
The Hamiltonian coefficients corresponding to five distinct bond
distances---employed in the present experimental implementation---are
tabulated in the Supplemental Material.
The VTC circuit $U(\boldsymbol{\theta})$ is composed of two layers of
hardware-efficient ansatz, each layer containing $R_Y$ and $R_Z$ rotations
for individual qubits, as well as a two-qubit CZ gate.
Further details including the explicit gate ordering and parameter indexing
are also provided in the Supplemental Material.

VTC time evolution is carried out at five bond
distances---$R = 0.25,\,0.75,\,1.25,\,1.75,\,2.25\,\mathrm{\AA}$---each
initialized to one of the computational basis states $\ket{00}$, $\ket{01}$,
or $\ket{11}$. To fulfill the Nyquist-Shannon sampling criterion, temporal
discretization is adapted to the spectral scale of the system: for the
shortest bond distance ($R = 0.25\,\mathrm{\AA}$), where the energy-level
spacing is largest, we set $N = 100$ time steps with step size $\tau = 0.1$,
yielding a total evolution time $T = 10\,\text{(a.u.)}$. For the remaining
bond distances ($R = 0.75$--$2.25\,\mathrm{\AA}$), where near-degeneracies
emerge and spectral resolution requirements increase, we adopt $N = 100$--$400$
and fixed $\tau = 0.5$, resulting in total evolution times $T = 50$--$200\,\text{(a.u.)}$.
This adaptive strategy ensures sufficient sampling density to
resolve closely spaced eigenvalues.

For each bond distance $R$ and initial state, the corresponding
VTC circuit is executed on the quantum processor at discrete time
steps, utilizing parameters pre-optimized via classical simulation.
Similar to the prior work~\cite{Sui2026_3DIsingPlatonic},
gradient-based optimization via experimental
measurements is conceptually viable, yet remains computationally prohibitive
due to excessive measurement overhead and sensitivity to accumulated noise.
Following QST, the reconstructed density matrix is purified to a pure
state, which is then phase-aligned globally to ensure a consistent
reference frame. Figure~\ref{fig3}(a) depicts the fidelity
distribution---histogram (left) and cumulative distribution function
(CDF, right)---between experimental and VTC-simulated states for
$\ket{00}$ and $\ket{01}$ at $R = 2.25\,\mathrm{\AA}$. The state
fidelity remains consistently above $99.4\%$ across all measured time
steps, underscoring the high experimental precision and robustness
of the VTC-based dynamical reconstruction.
The time-domain signal is calculated from the overlap
amplitude $S(t)=\bra{\psi_0}U(t)\ket{\psi_0}$, with
$\ket{\psi_0} \in \{\ket{00},\,\ket{01},\,\ket{11}\}$. To lift the
sign ambiguity in the extracted eigenenergies, we employ $\operatorname{Im}[S(t)]$
for Fourier-based spectral estimation. Figures~\ref{fig3}(b) display
$\operatorname{Im}[S(t)]$ for $R = 2.25\,\mathrm{\AA}$ with initial
states $\ket{00}$ and $\ket{01}$, respectively.
This signal $\operatorname{Im}[S(t)]$ is subsequently fitted to a multi-frequency
sine model incorporating both a constant
offset and phase parameters:
$f(t) = C + \sum_k A_k \sin(\omega_k t + \varphi_k)$, while the fitting
frequency and amplitude can be roughly initialized by a coarse Fourier analysis
of the experimental data.
The fitted results exhibit excellent agreement with the experimental
data, as shown in Fig.~\ref{fig3}(b).

Fourier analysis of the fitted survival amplitudes yields the eigenenergy
pairs $(-0.9399,-0.3729)\,E_h$ and $(-0.9294,-0.3866)\,E_h$ for the
$\ket{00}$ and $\ket{01}$ initial states, respectively (Fig.~\ref{fig3}(c)).
These values deviate from exact diagonalization (ED) of $H_{\mathrm{eff}}$
by less than $2\times10^{-4}\,E_h$. We apply this protocol consistently
across all five bond distances ($R = 0.25$--$2.25\,\mathrm{\AA}$).
As detailed in Table S2, comparisons between ED, numerical VTC,
and experimental results reveal a maximum absolute deviation
of $2\times10^{-3}\,E_h$ across the entire dataset.
Figure~\ref{fig3}(d) compares the experimentally
extracted energies of the four lowest-lying eigenstates versus bond distance
against the theoretical potential energy curve, demonstrating
quantitative agreement within the spectral resolution limit.
The dynamic fitting and Fourier transform analysis methods used here
are consistent in principle with those reported in Ref.~\cite{Sui2026_3DIsingPlatonic};
the difference lies in that this study is based on quantum dynamical
evolution data obtained from experiments, whereas prior research
relied on dynamical trajectories generated through classical
numerical simulations. Combining parameterized preparation with
VTC evolution imposes prohibitive circuit depth, rendering
practical demonstration contingent on future hardware advances.

Having demonstrated the efficacy of VTC in simulating the covalent-to-dissociative
transition in $\mathrm{H_2}$---a molecular prototype of electron
correlation---we now extend our investigation to the Fermi-Hubbard model and
the Mott metal-insulator transition (MIT) of this condensed matter system.
We employ dynamical eigenspectroscopy to systematically probe the spectral
evolution across this transition.
Via two-site dynamical mean-field theory (DMFT)
~\cite{Keen2020TwoSiteDMFT,Nie2024SelfConsistentSIAM},
the Fermi-Hubbard model maps onto a half-filled minimal Anderson
impurity model comprising one interacting impurity orbital
and one non-interacting bath orbital. Here, $U$ denotes the on-site Coulomb
repulsion at the impurity site, while $V$ characterizes the
hybridization (tunneling amplitude) between the impurity and
bath orbitals. Encoding the two spatial orbitals---each with
spin-up and spin-down degrees of freedom---into four qubits
yields the effective Hamiltonian:
\begin{equation}
\label{eq:hubbard-ham}
H = \frac{U}{4}\,Z_1 Z_3
  + \frac{V}{2}(X_1 X_2 + Y_1 Y_2 + X_3 X_4 + Y_3 Y_4),
\end{equation}
where qubits 1 and 3 represent the spin-up and spin-down modes of
the impurity orbital, and qubits 2 and 4 correspond to the spin-up and
spin-down modes of the bath orbital.
Although this minimal model is limited by finite-size effects and insufficient to
reproduce the true metal-insulator transition (MIT), its spectral characteristics
can still effectively capture the crossover behavior.
This mapping allows us to experimentally emulate three characteristic
regimes by tuning the interaction parameters~\cite{Nie2024SelfConsistentSIAM}:
a metallic state ($U=0$, $V=0.171$), an intermediate correlated
state ($U=1$, $V=0.121$), and the Mott insulating phase ($U=2$, $V=0.082$).
This Hamiltonian exhibits particle-hole symmetry due to the
half-filled condition, ensuring that the eigenspectrum is symmetric about zero energy.
Exploiting this particle-hole symmetry, we simplify the analysis
by reporting only the three distinct non-negative eigenenergies
$|E_k|$ ($k=0,1,2$) for each interaction regime.

\begin{figure*}[tp]
    \centering
    \includegraphics[width=0.8\linewidth]{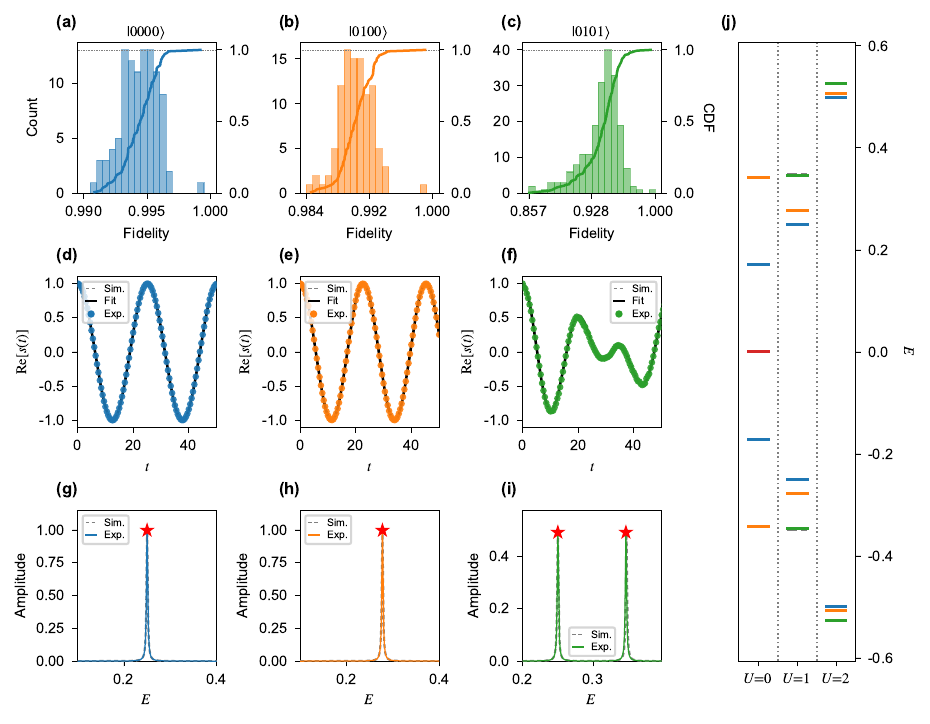}
    \caption{Experimental VTC dynamics and eigenspectra.
    (a)-(c) Fidelity histograms (left) and CDFs (right) comparing
    experimental states against VTC simulations for $\ket{0000}$,
    $\ket{0001}$, and $\ket{0101}$.
    (d)-(f) Experimental $\operatorname{Re}[S(t)]$ (Exp.) versus
    simulations (Sim., gray) and cosine fits (black) for the
    three initial states, demonstrating high fidelity across the evolution.
    (g)-(i) Fourier-transformed spectra revealing distinct eigenenergy peaks for each initial state.
    (j) Extracted eigenspectra (bars) for the three phases
    compared with exact diagonalization (gray dashed),
    confirming the accuracy of the VTC protocol.
    }
    \label{fig4}
\end{figure*}

Across all interaction regimes, we initialize the system in
$\ket{0000}$, $\ket{0001}$ or equivalent $\ket{0100}$,
and $\ket{0101}$ to extract signals and eigenenergies.
These states undergo coherent time evolution, followed
by full QST at discrete time steps. After post-measurement
purification and global phase alignment, we extract the
real part of the amplitude projected onto the initial state
to construct the time-domain signal $\operatorname{Re}[S(t)]$.
This signal is fitted to a multi-frequency cosine model,
$f(t) = C + \sum_k A_k \cos(\omega_k t + \varphi_k)$.
Owing to particle-hole symmetry, which enforces eigenenergy
pairing ($\pm E_k$), the signal exhibits
purely cosinusoidal behavior; thus, the fitted frequencies
$\omega_k$ correspond directly to the absolute eigenenergies
$|E_k|$. This unified, self-consistent protocol facilitates
the extraction of all eigenenergies, including
the ground-state energy $|E_0|$. As an independent benchmark,
ground-state energies across the phase diagram are also
computed using VQE;
detailed numerical results, convergence analysis,
and ansatz specifications are provided in the
Supplemental Material.

In the metallic phase, the four interaction terms in the Hamiltonian
$H = \tfrac{V}{2}(X_1 X_2 + Y_1 Y_2 + X_3 X_4 + Y_3 Y_4)$ commute pairwise.
This permits an exact factorization of the time-evolution operator,
thereby eliminating Trotter errors:
\begin{equation}\label{eq:metal-exact}
e^{-iHt}
= e^{-i\frac{V}{2}X_1 X_2 t}
e^{-i\frac{V}{2}Y_1 Y_2 t}
e^{-i\frac{V}{2}X_3 X_4 t}
e^{-i\frac{V}{2}Y_3 Y_4 t}.
\end{equation}
Accordingly, VTC is bypassed, enabling an exact, Trotter-error-free
implementation with a shallow circuit comprising 20 single-qubit
gates and 8 CNOTs. Initializing the system in $\ket{0000}$ and $\ket{0001}$,
we perform time-series evolution with step size $\tau = 1.0$,
sampling 101 time points up to $t = 100$. For the $\ket{0101}$ state,
we adopt a finer step $\tau = 0.5$, acquiring 101 points over $t \in [0, 50]$.
The average experimental-numerical state fidelity reached
$\bar{f} = 99.30\%$, with a minimum fidelity of $98.78\%$.
Multi-frequency cosine fitting and Fourier analysis
resolved three distinct eigenenergies $|E| = 0,\,0.1710,\,0.3423$,
exhibiting a maximum deviation of $3\times10^{-4}$ from theoretical predictions.

In the intermediate phase ($U=1$, $V=0.121$), the Hamiltonian terms
are generally non-commuting (e.g., $[Z_1Z_3,\,X_1X_2] \neq 0$).
Consequently, VTC compression is employed to mitigate the accumulation
of Trotter errors during long-time evolution. The VTC ansatz
features a hardware-efficient structure, with each layer comprising
4 $R_Y$ gates, 4 $R_Z$ gates, and 3 CNOT gates. We utilize two
layers ($L=2$, 16 parameters) for the initial states $\ket{0000}$
and $\ket{0100}$, while adopting three layers ($L=3$, 24 parameters)
for $\ket{0101}$ to enhance circuit expressibility. Numerical
simulations confirm robust convergence, as illustrated in
Fig.~\ref{fig2}(c,d).
Experimentally, we achieve average fidelities of
$\bar{f}_{\ket{0000}} = 99.44\%$, $\bar{f}_{\ket{0100}} = 99.02\%$,
and $\bar{f}_{\ket{0101}} = 93.86\%$. 
Experimentally extracted eigenenergies ($|E_0|=0.3467,$ $|E_1|=0.2776$, $|E_2|=0.2500$)
match numerical benchmarks with a maximum discrepancy of
$1.2\times10^{-3}$ in $|E_0|$.

In the Mott-insulating phase ($U=2$, $V=0.082$), the ratio
$U/V \approx 24 \gg 1$ indicates that the on-site interaction term
$\tfrac{U}{4}Z_1 Z_3$ overwhelmingly dominates the spectral structure.
In the limit $V \to 0$, the energy levels exhibit a threefold
degeneracy at $\pm U/4 = \pm 0.5$. Nonzero $V$ acts as a
perturbation, lifting this degeneracy and splitting the levels
into three closely spaced multiplets: $\pm 0.5000,\,\pm 0.5067,\,\pm 0.5262$.
The VTC protocol retains the same hardware-efficient ansatz
structure as employed in the intermediate phase: two layers ($L=2$)
for initial states $\ket{0000}$ and $\ket{0001}$, and three
layers ($L=3$) for $\ket{0101}$. Long-time VTC evolution, analyzed
via multi-frequency cosine fitting, fully resolves these near-degenerate
levels. The experimentally extracted energies are
$|E_0| = 0.5266$, $|E_1| = 0.5066$, and $|E_2| = 0.4993$,
with a maximum deviation of $0.0007$ from the theoretical predictions.
For the two $L=2$ trajectories initialized in $\ket{0000}$ and
$\ket{0001}$, the experimental state fidelities range from $98.88\%$ to
$99.93\%$, with an average of $99.38\%$. By contrast, the deeper
$L=3$ trajectory initialized in $\ket{0101}$ achieves a maximum
experimental state fidelity of $99.61\%$, with an average of $95.25\%$.

Figure~\ref{fig4} summarizes the core experimental results across
the three phase regions. Focusing on the intermediate phase
($U=1$): Figure~\ref{fig4}(a)-\ref{fig4}(c) illustrate the
fidelity trajectories between experimental outcomes and VTC-simulated
states for $\ket{0000}$, $\ket{0100}$, and $\ket{0101}$, yielding
average fidelities of $99.44\%$, $99.02\%$, and $93.86\%$, respectively.
The slight reduction in fidelity for $\ket{0101}$ stems from the
deeper $L=3$ ansatz, which is more susceptible to experimental
noise than its $L=2$ counterpart. Figure~\ref{fig4}(d)-\ref{fig4}(f)
corroborate the strong agreement between the experimental and
simulated time-domain signals $\operatorname{Re}[S(t)]$, with
multi-frequency cosine fits accurately capturing the dynamics.
Figure~\ref{fig4}(g)-\ref{fig4}(i) display the eigenenergy
spectra extracted via FFT of these fitted signals. For $\ket{0000}$,
a single peak is resolved at $E=0.2500$, consistent with the
fact that this state coincides with the Hamiltonian's eigenstate
at $E=0.25$. For $\ket{0100}$, although a single peak is observed
experimentally ($E=0.2776$), spectral analysis reveals it
originates from the superposition of eigenstates at $\pm 0.2777$.
For $\ket{0101}$, two distinct peaks are extracted ($E=0.2500$ and
$0.3467$), aligning with the simulated composition of eigenstates
at 0.2500 and $0.3479$; the marginally larger deviation here is
attributable to the increased circuit depth. Finally, Fig.~\ref{fig4}(j)
consolidates the complete energy-level diagrams for all phases.
Despite the inherently discrete nature of the four-qubit spectrum,
the crossover from the metallic state (featuring a zero-energy Fermi level)
to the gapped Mott insulator emerges as a distinct transition.

\section{Discussion and Conclusion}

In summary, we have successfully implemented a high-resolution
experimental protocol for extracting the eigenspectra of correlated
quantum systems, specifically demonstrating the approach on a
four-qubit Fermi-Hubbard simulator. The central achievement of
this work lies not merely in the application of a specific
algorithm, but in the establishment of a complete experimental
pipeline: the integration of controlled time evolution,
post-measurement purification, global phase alignment,
and Fourier-based spectral analysis.

While VTC proved instrumental in mitigating Trotter errors
and enabling access to longer effective evolution times beyond the native
coherence limits of the hardware, it functioned primarily
as a facilitator for the overarching goal of precision
spectroscopy. By maintaining evolved state fidelities
above $99\%$ over extended time scales, VTC provided
the dense, coherent time-series data required to resolve
energy splittings as small as $10^{-3}$. This capability
was critical for disentangling the near-degenerate multiplets
characteristic of the Mott insulating phase, a task
infeasible with standard short-depth Trotterization
due to rapid signal decoherence.

Experimentally, we achieved high-fidelity state
preparation and accurate eigenenergy
extraction across all tested regimes. The slight fidelity
reduction observed for deeper circuits (e.g., $L=3$ ansatz for
$\ket{0101}$) underscores the ongoing challenge of balancing
expressibility with decoherence---a trade-off that future
error mitigation techniques could address. The consistency
between VTC-extracted energies and exact diagonalization
validates our hybrid approach of combining experimental
evolution with classical post-processing (purification
and phase alignment), which circumvents the need for
ancilla qubits in phase-sensitive measurements.

Extending this approach to larger systems entails challenges 
common to parameterized quantum circuits--namely, 
balancing ansatz expressibility with optimization trainability. 
Our results demonstrate that circuit-depth scanning 
effectively enhances the ansatz's representational 
capacity. For the barren plateau problem, our 
protocol also partially alleviates it through a 
built-in warm-start mechanism: parameters optimized 
at the previous Trotter step initialize the next, 
keeping the variational search in a favorable parameter 
regime. Another bottleneck of scaling is the exponential 
overhead of QST measurement. Adopting informationally 
complete measurements---such as SIC-POVMs or shallow 
tomography---can reduce the number of
experimental shots required for 
tomography~\cite{Hu2024SampleEfficientQST,You2025CircuitOptimizationICPOVM}.

Beyond these specific implementations, VTC's ability 
to resolve both ground and excited states via Fourier 
analysis of time-domain survival amplitudes establishes 
a unified spectroscopic framework that goes beyond 
the ground-state-centric paradigm of conventional VQE. 
The seamless transition demonstrated here---from the 
$\mathrm{H_2}$ molecule in quantum chemistry to the 
Hubbard model in condensed matter---highlights VTC's 
versatility in probing correlation-driven phenomena, 
including the Mott crossover. As quantum processors 
scale toward larger qubit counts and improved fidelities, 
this hybrid dynamics-compression approach provides 
a practical, hardware-efficient route to spectral 
reconstruction in regimes that are intractable 
for traditional Trotter methods.

\begin{acknowledgments}
The authors from Zhejiang University (Y.Y. and colleagues) thank the
support from the National Natural Science Foundation of China (Grant No. 12074336).
\end{acknowledgments}

\section*{DATA AVAILABILITY}

The data supporting the findings of this study are available from the Zenodo
repository at \url{https://doi.org/10.5281/zenodo.22020779}~\cite{Sui2026VTCDataset}.

\appendix

\section{Readout Correction, Quantum State Tomography, and State Purification}
\label{sec:qst-readout}

The readout fidelities for the ground state $\ket{0}$ and excited
state $\ket{1}$ of the qubits are approximately $98.8\%-99.7\%$ and
$92.8\%-98.4\%$, respectively. To compensate for readout errors,
a pre-calibrated assignment error correction matrix $\mathbf{F}$
is introduced, applying the inverse transformation
$\boldsymbol{p}^{\mathrm{corr}} = \mathbf{F}^{-1}\boldsymbol{p}^{\mathrm{raw}}$
to the raw measurement population probabilities, thereby
improving the accuracy of density matrix reconstruction. When necessary,
we recalibrate the readout correction matrix at each time step to
track system drift.

After the circuit execution, the density matrix $\rho$ is
reconstructed via quantum state tomography (QST)~\cite{NielsenChuang2010}.
An $n$-qubit density matrix can be expanded in the Pauli
operator basis $\{\mathcal{O}_\alpha\}$:
\begin{equation}\label{eq:dm-decomp}
\rho = \frac{1}{2^n}\sum_\alpha \rho_\alpha\,\mathcal{O}_\alpha,
\quad
\rho_\alpha = \mathrm{tr}(\mathcal{O}_\alpha\rho),
\end{equation}
Each coefficient $\rho_\alpha$ is extracted by measuring the
population distribution after rotating the quantum state into
the corresponding Pauli basis.
For the two-qubit $\mathrm{H_2}$ system, we employ 9 measurement settings
$\{I,\,-Y/2,\,X/2\}^{\otimes 2}$, each followed by computational-basis
readout (5000 shots). These settings suffice to reconstruct all $4^2=16$
Pauli coefficients $\rho_\alpha = \mathrm{tr}(\mathcal{O}_\alpha\rho)$
in Eq.~\eqref{eq:dm-decomp}---the $Z$ components are accessed directly
from the $I$-setting computational readout, while $X$ and $Y$ components are
accessed via the $-Y/2$ and $X/2$ single-qubit rotations---fully determining
the $4\times4$ density matrix. Correspondingly, for the four-qubit Hubbard
system, we employ 81 measurement bases from the set
$\{I,\,-Y/2,\,X/2\}^{\otimes 4}$ to fully reconstruct the
$16\times16$ density matrix from 256 Pauli coefficients.

Owing to finite sampling noise and hardware imperfections,
the raw density matrix $\rho^{\exp}$ reconstructed via
Eq.~\eqref{eq:dm-decomp} often violates physicality
constraints---specifically, positive semidefiniteness and
unit trace. To obtain a physically consistent pure-state estimate for time-domain
analysis, we apply a maximum-likelihood-inspired physicality projection
to the experimentally reconstructed density matrix $\rho^{\exp}(t)$.
We first diagonalize it as
\begin{equation}
\rho^{\exp}(t)=\sum_j \lambda_j(t)\ket{v_j(t)}\bra{v_j(t)}.
\end{equation}
The negative eigenvalues are removed by defining
$\lambda_j^{+}(t)=\max[\lambda_j(t),0]$. The remaining eigenvalues are
then renormalized according to
$\widetilde{\lambda}_j(t)=\lambda_j^{+}(t)/\mathcal{N}_{+}(t)$, where
$\mathcal{N}_{+}(t)=\sum_k\lambda_k^{+}(t)>0$. The resulting
positive-semidefinite density matrix is
\begin{equation}
\label{eq:rho-psd}
\rho_{\mathrm{PSD}}^{\exp}(t)=\sum_j\widetilde{\lambda}_j(t)\ket{v_j(t)}\bra{v_j(t)}.
\end{equation}
By construction, the normalized eigenvalues satisfy
$\widetilde{\lambda}_j(t)\geq0$ and
$\sum_j\widetilde{\lambda}_j(t)=1$, so that
$\rho_{\mathrm{PSD}}^{\exp}(t)$ is a trace-one positive-semidefinite
density matrix. To obtain a pure-state estimate, we order the normalized
eigenvalues as
$\widetilde{\lambda}_1(t)\geq\widetilde{\lambda}_2(t)\geq\cdots\geq0$
and take the eigenvector associated with the largest eigenvalue:
\begin{equation}
\label{eq:purification}
\ket{\psi^{\exp}(t)}=\ket{v_1(t)}.
\end{equation}

The dominant eigenvector of $\rho_{\mathrm{PSD}}^{\exp}(t)$ defines the
optimal pure-state, or rank-one-projector, approximation to
$\rho_{\mathrm{PSD}}^{\exp}(t)$ in the Frobenius norm. Specifically, the
projector $\ket{v_1(t)}\bra{v_1(t)}$ minimizes
$\left\|\rho_{\mathrm{PSD}}^{\exp}(t)-\ket{\psi}\bra{\psi}\right\|_F$
over all normalized pure states $\ket{\psi}$. When the reconstructed
state is close to a pure state, the dominant eigenvalue
$\widetilde{\lambda}_1(t)$ remains close to unity, while the subdominant
eigenvalues remain small. The PSD projection removes negative-eigenvalue
artifacts introduced by the density-matrix reconstruction, whereas the
subsequent dominant-eigenvector extraction retains the principal coherent
component for time-domain analysis.

\section{Global-Phase Alignment in QST Post-Processing}
\label{sec:phase}
QST reconstructs the experimental density matrix $\rho^{\exp}$,
which is subsequently purified via eigen-decomposition to yield
the dominant experimental state vector $\ket{\psi^{\exp}(t)}$.
For any normalized pure state $\ket{\psi}$, multiplication
by a global phase factor $e^{i\Delta\varphi}$ leaves the associated
density matrix invariant:
\begin{equation}\label{eq:dm}
\rho' = \bigl(e^{i\Delta\varphi}\ket{\psi}\bigr)
\bigl(e^{i\Delta\varphi}\ket{\psi}\bigr)^\dagger
= \ket{\psi}\bra{\psi} = \rho.
\end{equation}
Consequently, the experimentally reconstructed state vector
$\ket{\psi^{\exp}(t)}$ exhibits an arbitrary global phase
ambiguity relative to the numerically simulated reference state.
This phase indeterminacy prevents direct extraction of
eigenenergy values from the time-domain signal via Fourier
analysis---since spectral peaks in the power spectrum depend
on relative phase coherence across time steps.

In principle, given the experimentally reconstructed state
vector $\ket{\psi^{\exp}(t)}$, the time evolution of the
expectation value $\langle A\rangle(t)=\bra{\psi^{\exp}(t)} A\ket{\psi^{\exp}(t)}$ for a suitably
chosen observable $A$ (like Pauli string operator) can be directly calculated as
\begin{equation}\label{eq:operator}
\langle A\rangle(t) = \sum_{m,n} c^*_m c_n\, A_{mn}\, e^{i(E_m - E_n)t},
\end{equation}
where $A_{mn} = \bra{\phi_m} A \ket{\phi_n}$ denotes the matrix
element of $A$ in the energy eigenbasis $\{ \ket{\phi_m}, \ket{\phi_n} \}$,
and the coefficients $c_m = \langle \phi_m|\psi_0\rangle$
encode the initial-state decomposition.
For a Hermitian observable $A$, $\langle A\rangle(t)$ is real and
invariant under a global phase transformation of the quantum state.
Fourier analysis of this time-domain signal yields spectral
peaks at frequencies $\omega_{mn} = E_m - E_n$, thereby
resolving pairwise energy differences.
A constrained combinatorial optimization algorithm can be
applied to the energy differences to reconstruct an
ordered set of absolute eigenenergies $E_0 < E_1 < E_2 < \cdots$,
where the ground-state energy $E_0$ can be independently determined
by VQE.

Here for simplicity, we perform further data post-processing on
the experimental state vector $\ket{\psi^{\exp}(t)}$ using
phase alignment assisted by numerical simulation to
resolve the global phase issue, as follows.
The relation between the experimental state and the
theoretical ideal can be expressed as
\begin{equation}\label{eq:phase-diff}
\ket{\psi^{\exp}(t)}
= e^{i\Delta\varphi(t)}\ket{\psi^{\mathrm{sim}}(t)}
+ \ket{\delta\psi(t)},
\end{equation}
where $\ket{\delta\psi(t)}$ is the residual term arising from measurement noise and finite fidelity.
Denote the components of the two states as
\begin{equation}\label{eq:states}
\ket{\psi^{\mathrm{sim}}(t)} =
\begin{pmatrix} A_1 e^{i\varphi_1} \\ A_2 e^{i\varphi_2} \\ \vdots \end{pmatrix},
\quad
\ket{\psi^{\exp}(t)} =
\begin{pmatrix} A_1' e^{i\varphi_1'} \\ A_2' e^{i\varphi_2'} \\ \vdots \end{pmatrix},
\end{equation}
where $A_j, A_j' \geq 0$ are the amplitudes, and
$\varphi_j, \varphi_j'$ are the corresponding phases.

Select a component with larger amplitude (higher phase measurement
signal-to-noise ratio) as the reference, for example, choose the first
component with $j=1$, and multiply the experimental state by the correction
factor $e^{i(\varphi_1 - \varphi_1')}$ ($\varphi_1$ obtained from
numerical simulation, $\varphi_1'$ extracted from QST measurements),
to obtain the phase-aligned state.
\begin{equation}\label{eq:corrected-state}
\ket{\psi^{\mathrm{cor}}(t)}
= e^{i(\varphi_1-\varphi_1')}
\begin{pmatrix} A_1' e^{i\varphi_1'} \\ A_2' e^{i\varphi_2'} \\ \vdots \end{pmatrix}
=
\begin{pmatrix} A_1' e^{i\varphi_1} \\ A_2' e^{i(\varphi_2'+\Delta\varphi)} \\ \vdots \end{pmatrix},
\end{equation}
where $\Delta\varphi \equiv \varphi_1 - \varphi_1'$.
This operation only rotates the global phase and does not alter
population probabilities in any measurement basis.
For simple computational basis initial states such as $\ket{00}$,
after phase alignment, the first component $A_1' e^{i\varphi_1}$ of
$\ket{\psi^{\mathrm{cor}}(t)}$ serves as the experimental estimate
of the time-domain signal $\bra{00}U(t)\ket{00}$.
For more complex initial states $\ket{\psi_0}$,
$\ket{\psi^{\mathrm{cor}}(t)}$ can be used to compute the inner
product with the initial state, yielding the time-domain
signal $\bra{\psi_0}U(t)\ket{\psi_0}$. This phase-alignment
protocol follows our prior work on real-valued Hamiltonians:
we previously fixed eigenstate signs via numerical validation
of measured populations~\cite{Sui2026_3DIsingPlatonic},
and now supplement population data with simulated phases,
as direct global phase readout is infeasible on our hardware.

\nocite{*}

\bibliography{Ref}

\end{document}